\documentclass[aps,twocolumn,nofootinbib,preprintnumbers]{revtex4}

\usepackage[dvips]{color}
\usepackage[normalem]{ulem}
\usepackage{amsmath}
\usepackage{enumerate}
\usepackage{amsfonts}
\usepackage{epsfig}

\usepackage{graphicx}
\usepackage{bm}

\def\eg{{\it e.g. }}
\def\ie{{\it i.e.}}

\def\({\left(}
\def\){\right)}
\def\[{\left[}
\def\]{\right]}
\def\<{\langle}
\def\>{\rangle}

\newcommand\half{{\ensuremath{\frac{1}{2}}}}
\newcommand\p{\ensuremath{\partial}}

\newcommand\field[1]{{\ensuremath{\mathbb{{#1}}}}}

\newcommand\vev[1]{{\ensuremath{\left\langle{#1}\right\rangle}}}

\newcommand{\RR}{\field{R}}

\newcommand{\ZZ}{\field{Z}}

\newcommand{\be}{\begin{equation}}
\newcommand{\ee}{\end{equation}}
\newcommand{\bea}{\begin{eqnarray}}
\newcommand{\eea}{\end{eqnarray}}

\newcommand{\bi}{\begin{itemize}}
\newcommand{\ei}{\end{itemize}}
\newcommand{\ben}{\begin{enumerate}}
\newcommand{\een}{\end{enumerate}}
\newcommand{\bca}{\begin{cases}}
\newcommand{\eca}{\end{cases}}
\newcommand{\bln}{\begin{align}}
\newcommand{\eln}{\end{align}}
\newcommand{\bst}{\begin{split}}
\newcommand{\est}{\end{split}}

\newcommand\al{{\alpha}}
\newcommand\ep{\epsilon}
\newcommand\sig{\sigma}

\newcommand\lam{\lambda}

\newcommand\om{\omega}

\newcommand\ga{{\ensuremath{{\gamma}}}}
\newcommand\Ga{{\ensuremath{{\Gamma}}}}

\newcommand\De{{\ensuremath{{\Delta}}}}

\newcommand\da{{\dagger}}

\newcommand\ov{\over}
\newcommand\ha{{\half}}

\def\le{\left}
\def\ri{\right}

\newcommand\sD{{\ensuremath{{\mathcal D}}}}

\newcommand\sN{{\ensuremath{{\mathcal N}}}}
\newcommand\sO{{\ensuremath{{\mathcal O}}}}
\newcommand\sR{{\ensuremath{{\mathcal R}}}}

\newcommand\ut{{\underline{t}}}

\newcommand\sS{{\mathcal S}}

\newcommand\bpsi{{\bar \psi}}

\renewcommand{\Im}{\textrm{Im}\,}
\renewcommand{\Re}{\textrm{Re}\,}

\begin{document}

\title{Non-Fermi liquids from holography}

\preprint{MIT-CTP/4023}

\author{Hong Liu, John McGreevy and David Vegh}

\affiliation{Center for Theoretical Physics,
Massachusetts
Institute of Technology,
Cambridge, MA 02139
}

%

\begin{abstract}

We report on a potentially new class of non-Fermi liquids in (2+1)-dimensions. They are
identified via the response functions of composite fermionic operators in a class of strongly
interacting  quantum field theories at finite density, computed using the AdS/CFT
correspondence. We find strong evidence of Fermi surfaces: gapless fermionic excitations at
discrete shells in momentum space. The spectral weight exhibits novel phenomena, including
particle-hole asymmetry, discrete scale invariance, and scaling behavior consistent with that of
a critical Fermi surface postulated by Senthil.

\end{abstract}

\maketitle

\section{Introduction}

\nocite{anderson} \nocite{varma} \nocite{varmarev}

The normal state of the high-$T_C$ superconducting cuprates, and metals close to a quantum
critical point are examples of non-Fermi liquids, which have sharp Fermi surfaces but whose low
energy properties differ significantly from those predicted from Landau's Fermi liquid
theory~\cite{anderson, varma, varmarev}. While Landau Fermi liquids are controlled by a free
Fermi gas fixed point with (almost) no relevant perturbations~\cite{Benfatto, ShankarA,
Polchinski:1992ed, Shankar:1993pf}, a proper theoretical framework characterizing non-Fermi
liquid metals, which likely involves strong couplings, is lacking. In this paper we search for
new universality classes of non-Fermi liquids using the AdS/CFT
correspondence~\cite{Maldacena:1997re, Gubser:1998bc, Witten:1998zw}.

According to the correspondence, any (quantum) gravity theory in a $(d+1)$-dimensional
asymptotically anti-de Sitter (AdS$_{d+1}$) spacetime is dual to a $d$-dimensional quantum field
theory ``living at its boundary''. Through the AdS/CFT dictionary, a gravity theory can (in
principle) be used to obtain all physical observables of its boundary dual, like the physical
spectrum and correlation functions. Compared to conventional approaches, the gravity approach
offers some remarkable features which make it a valuable tool for discovering new strongly
coupled phenomena:

\ben

\item At small curvature and low energies known gravity theories reduce to a universal sector:
classical Einstein gravity plus matter fields. Through the duality, this limit typically translates into the strong-coupling and large-$N$ limit
of the boundary theory, where $N$ characterizes the number of species. Thus by working with
Einstein gravity (plus various matter fields) one can extract certain universal properties of a
large number of {\it strongly coupled} quantum field theories\footnote{The field theory origin
of such universality is still rather mysterious.}.

\item Highly dynamical, strong-coupling phenomena in the dual field theories
can often be understood on the gravity side using simple geometric pictures. Familiar examples
include confinement and chiral symmetry breaking in a non-Abelian gauge theory.

\item Putting the boundary theory at finite temperature and finite density corresponds to putting a black hole in the bulk geometry.
Questions about complicated many-body phenomena at strong coupling can be answered by solving
linear wave equations in this black hole background.

\een

Consider a quantum field theory which contains fermions charged under a global $U(1)$
symmetry\footnote{The theory may also contain charged scalars, and both scalars and fermions may
couple additionally to some non-Abelian gauge bosons.}. When a finite $U(1)$ charge density is
introduced into such a theory, it is natural to ask whether the system possesses a Fermi surface
and if yes, what are the low-energy excitations. One approach to these questions is to study
spectral functions of a fermionic composite operator\footnote{The spectral function of an
operator is a measure of the density of states which couple to the operator. It is proportional
to the imaginary part of the retarded function $G_R$ of the operator in momentum space.}. The
presence of a Fermi surface may be revealed by the appearance of gapless excitations of the
operator at discrete shells in momentum space.

This will be the approach taken here.  We obtain the spectral functions for fermionic operators
using the classical Einstein gravity. We will work in $(2+1)$ dimensions and leave
other-dimensional theories for future study. While the string theory landscape in principle
provides a large number of AdS/CFT dual pairs, only a few explicit examples are understood in
detail; in $(2+1)$-dimensions, these include the $\sN=8$ M2-theory and recently-discovered ABJM
theories~\cite{Bagger:2007vi, Gustavsson:2007vu, Aharony:2008ug}. As emphasized earlier,
however, by working with the classical Einstein gravity we are extracting universal properties
of a large number of boundary field theories, even if their explicit Hamiltonians are not known.
The spectral functions we find give strong indications of the presence of Fermi surfaces of some
non-Fermi liquid. We find poles representing `marginal' quasi-particles at discrete shells in
momentum space, with scaling behavior different from that of a Landau quasi-particle. We also
observe some other novel phenomena, including particle-hole asymmetry and discrete scale
invariance for a continuous range of momenta.

Our investigation was motivated in part by earlier work of Sung-Sik Lee~\cite{Lee:2008xf}, which
initiated the study of spectral functions of fermionic operators using a gravity dual. Our
results differ from those of~\cite{Lee:2008xf}; we believe the difference lies in the
implementation of the real-time holographic prescription \cite{Son:2002sd,Iqbal:2008by, iqbalN}.

The plan of the paper is as follows. In the next section we set up the framework for calculating
the spectral functions of a fermionic operator at finite density using the gravity description.
In sec.~\ref{sec:3} we discuss properties of the spectral functions, including scaling behavior
near a Fermi surface. We conclude in sec.~\ref{sec:4} with a discussion of the interpretation of
the results and possible caveats.

\section{Set-up of the calculation} \label{sec:2}

Consider a three-dimensional relativistic conformal field theory (CFT) with a global $U(1)$
symmetry that has a gravity dual. Such a system at finite charge density can be described by a
charged black hole in four dimensional anti-de Sitter spacetime
(AdS$_4$)~\cite{Chamblin:1999tk}, with the current $J_\mu$ in the CFT mapped to a $U(1)$ gauge
field $A_M$ in AdS. A fermionic operator $\sO$ in the CFT with charge $q$ and conformal
dimension $\De$ is mapped to the gravity side to a spinor field $\psi$ with charge $q$ and a
mass
\be
m R = \De -{3 \ov 2}
\ee
where $R$ is the AdS curvature radius. The spectral function of $\sO$ at finite charge density
can then be extracted by solving the Dirac equation for $\psi$ in the charged AdS black hole
geometry. Which pairs of $(q, \De)$ arise depends on the specific dual CFT. However, since we
are working with a universal sector common to many gravity theories, we will take the liberty of
considering an arbitrary pair of $(q, \De)$, scanning many possible CFTs.

\subsection{Black hole geometry}

The action for a vector field $A_M$ coupled to AdS$_4$ gravity can be written as
 \be \label{grac}
 S = {1 \ov 2 \kappa^2} \int d^{4} x \,
 \sqrt{-g} \le[\sR - {6 \ov R^2} - {R^2 \ov g_F^2} F_{MN} F^{MN} \ri]
 \ee
where $g_F^2$ is an effective dimensionless gauge coupling\footnote{It is defined so that for a
typical supergravity Lagrangian it is a constant of order $O(1)$}. The equations of motion
following from~\eqref{grac} are solved by the geometry of a charged black
hole~\cite{Romans:1991nq, Chamblin:1999tk}\footnote{For a generic embedding of (\ref{grac}) into
4d ${\cal N}=2$ supergravity, this solution can be lifted to an M-theory solution
\cite{Gauntlett:2007ma}.}
\be
ds^2  =  {r^2 \ov R^2} (-f dt^2 + d x_i^2)  +  {R^2 dr^2 \ov r^2 f}
\ee
with
\be
f = 1 + { Q^2 \ov r^{4}} - {M \ov r^3}, \quad A_0 = \mu \le(1- {r_0 \ov  r}\ri), \quad
\mu  \equiv  {g_F Q \ov R^2 r_0}
\ee
where $r_0$ is the horizon radius determined by $f(r_0) =0$, and $\mu $ can be identified as the chemical potential of the boundary theory. For calculational purposes it is convenient to use dimensionless quantities. Consider the rescaling
 \bea
&& r \to r_0 r, \quad (t,\vec x) \to {R^2 \ov r_0} (t , \vec x), \quad A_0 \to {r_0 \ov R^2} A_0, \nonumber \\
&& \qquad\quad M \to M r_0^3, \quad Q \to Q r_0^{2}
\eea
after which the metric becomes
\be \label{bhmetric}
{ds^2 \ov R^2} \equiv g_{MN} dx^M dx^N = r^2  (-f dt^2 + d\vec x^2)  + {1 \ov r^2} {dr^2 \ov f}, \qquad
\ee
with now the horizon at $r=1$ and
\be \label{eep}
f = 1 +{Q^2 \ov r^{4}} - {1+Q^2 \ov r^3}, \quad A_0 = \mu \le(1- {1 \ov r} \ri), \quad
\mu  = g_F Q  \ .
\ee
The dimensionless temperature is given by
 \be  \label{r10}
 T = {1 \ov 4 \pi} \le(3-Q^2 \ri) \ .
\ee
The zero-temperature limit is obtained by setting $Q =  \sqrt{3}$.
At zero temperature, near the horizon the metric~\eqref{bhmetric} becomes $AdS_2 \times \RR^{2}$ with the curvature radius of $AdS_2$ given by
 \be \label{epeo}
 R_2 = {R \ov \sqrt{6}} \ .
 \ee

\subsection{Dirac equation}

To compute the spectral functions for $\sO$ we need only  the
quadratic action of $\psi$ in the geometry~\eqref{bhmetric}-\eqref{epeo}
  \be \label{Dact}
 S_{\rm spinor} =  \int d^{d+1} x \sqrt{-g} \, i (\bar \psi \Ga^M \sD_M \psi  - m \bpsi \psi)
 \ee
where
 \be
\bpsi = \psi^\da \Ga^\ut, \quad \sD_M  = \p_M + {1 \ov 4} \om_{ab M} \Ga^{ab} - i q A_M
 \ee
and $\om_{ab M}$ is the spin connection\footnote{We will use $M$ and $a,b$ to denote bulk
spacetime and tangent space indices respectively, and $\mu, \nu \cdots$ to denote indices along
the boundary directions, i.e. $M = (r, \mu)$. All indices on Gamma matrices refer to tangent
space ones. For notational convenience below we will take $m$ to be defined in units of $1/R$,
i.e. $m R \to m$.}. Note that the Dirac action~\eqref{Dact} depends on $q$ only through
 \be \label{mme}
 \mu_q \equiv \mu q = g_F q Q
 \ee
\ie\ through the combination of $g_F q$. This is expected; $\mu$ is the minimal amount energy
needed to add a unit charge to the system, thus for a field of charge $q$, the effective
chemical potential is given by $\mu_q$. Below, for notational simplicity, we will set $g_F = 1$
and treat $q$ as a free parameter, but one should keep in mind only the product of them is the
relevant quantity.

To analyze the Dirac equations following from~\eqref{Dact}, it is convenient to use the following basis
\be \label{gct}
\Ga^r = \left( \begin{array}{cc}
1 & 0  \\
0 & -1
\end{array} \right), \;\; \Ga^\mu = \left( \begin{array}{cc}
0 & \ga^\mu  \\
\ga^\mu & 0
\end{array} \right), \;\; \psi = \( \begin{array}{c} \psi_+ \\ \psi_- \end{array}\)
\ee
where $\psi_\pm$ are two-component spinors and $\ga^{\mu}$ are gamma matrices of the (2+1)-dimensional boundary theory.
Writing
\be
\psi_\pm =  (- g g^{rr})^{-{1 \ov 4}} e^{-i \om t + i k_i x^i} \phi_\pm \ ,
 \ee
the Dirac equation becomes
\be \label{eropi}
\sqrt{ g_{ii} \ov  g_{rr}} \le( \p_r  \mp m \sqrt{g_{rr}}\ri) \phi_\pm
= \mp i K_\mu \ga^\mu \phi_\mp,
\ee
with
\be \label{uen}
K_\mu (r) = \(-u(r),  k_i \), \qquad  u= \sqrt{g_{ii} \ov - g_{tt}} (\om + \mu_q (1-{1 \ov r})) \ .
\ee
Note that since as $r \to \infty$, $u \to \om + \mu_q$,
$\om$ should correspond to the difference of the boundary theory frequency from $\mu_q$,
\ie\ $\om =0$ corresponds to the Fermi energy.

To extract the retarded Green function for $\sO$, we need to
solve~\eqref{eropi} with the in-falling boundary condition at the horizon~\cite{Son:2002sd}, and to identify the source and
the expectation value for $\sO$ from the asymptotic behavior of $\psi$ near the boundary. Such an identification
can be carried out from the prescription of~\cite{Iqbal:2008by, iqbalN}, which amounts to
identifying $\psi_+$ as the source and its canonical momentum in terms of $r$-slicing (which is essentially $\psi_-$)
as the expectation value.
 More explicitly, $\phi_\pm$ have the following asymptotic behavior near $r \to \infty$,
 \be \label{r0}
\phi_+ = A r^{ m} + B r^{- m -1}, \;
\phi_- = C r^{ m -1} + D r^{ - m } \
\ee
with
\be
C = {i \ga^\mu k_\mu \ov 2m-1} A, \; B = {i \ga^\mu k_\mu \ov 2m+1} D, \; k_\mu = (-(\om + \mu_q), k_i) \ .
\ee
Suppose the coefficients $D$ (corresponding to expectation value) and $A$ (corresponding to source) are
related by $D = \sS A$, then the retarded Green function $G_R$ is given
by~\cite{iqbalN}\footnote{Here we assume $m \geq 0$. For $m<0$, one simply exchanges the roles of
$A$ and $D$. For $m \in [0,\ha)$, both quantization procedures are allowed. Also note that the
factor $\ga^0$ in~\eqref{alfo} comes from $G_R \sim \vev{\{\sO, \sO^\da\}}$ while in perturbing
the boundary action we add $- i \int d^3 x \, (\bar \psi_0 \sO + \bar \sO \psi_0)$ with $\bar \sO
= \sO^\da \ga^0$.
As noted in \cite{iqbalN},
we must choose the sign of the overall gravity action to be consistent with unitarity.}
 \be \label{alfo}
 G_R = -i \sS \ga^0 \ .
 \ee

Equations~\eqref{eropi} can be further simplified by choosing the basis $\ga^0 = i \sig_2, \ga^1
= \sig_1,  \gamma^2 = \sig_3$ and setting $k_2 =0$,\footnote{Since the system is rotationally
symmetric, there is no loss of generality.} after which one finds two sets of decoupled
equations
\bea \label{set1}
&&\sqrt{ g_{ii} \ov  g_{rr}} \le( \p_r  \mp m \sqrt{g_{rr}}\ri)  y_\pm = \mp i (k_1 -u) z_\mp, \\
&&\sqrt{ g_{ii} \ov  g_{rr}} \le( \p_r  \pm m \sqrt{g_{rr}}\ri)  z_\mp =  \pm i (k_1 +u) y_\pm
\label{set2}
\eea
where we have written $\phi_\pm = \left( \begin{array}{c}
y_\pm  \\
z_\pm
\end{array} \right)
$. Introducing the ratios
\be
\xi_+ = {i y_- \ov  z_+} , \qquad \xi_- = - {i z_- \ov y_+},
\ee
and using~\eqref{alfo}, the retarded Green function $G_R$ can be written as
\be \label{bdG}
G_R = \lim_{\ep \to 0}  \ep^{-2m}  \left( \begin{array}{cc}
\xi_+ & 0  \\
0  & \xi_-
\end{array} \right) \biggr|_{r = {1 \ov \ep}},
\ee
where one should extract the finite terms in the limit.

It is convenient to derive flow equations directly for $\xi_{\pm}$
as in \cite{Iqbal:2008by}. From~\eqref{set1}, we find
\be \label{sor1}
 \sqrt{g_{ii} \ov g_{rr}} \p_r \xi_\pm = -2m \sqrt{g_{ii}} \xi_\pm \mp (k_1 \mp u)
\pm (k_1 \pm u) \xi_\pm^2
\ee
The in-falling boundary condition at the horizon implies
\be \label{tye}
\xi_\pm|_{r=1} =i \ .
\ee
With the boundary condition~\eqref{tye}, one can now integrate~\eqref{sor1} to $r \to \infty$ to
obtain the boundary correlation function directly. Below we will drop the subscript $1$ on
$k_1$.

At zero temperature, the $\omega \to 0$ limit of equations~\eqref{set1}--\eqref{set2}
and~\eqref{sor1} is singular, since $g_{tt}$ then has a double zero at the horizon. As we will
see this has important consequences for the behavior of $G_R$ near $\om =0$. Also, at $\om =0$
the in-falling boundary conditions~\eqref{tye} do not apply and should be replaced by
\be \label{roep}
 \xi_{\pm}|_{r=1, \, \om =0}= {  m - \sqrt{k^2 + m^2 - {\mu_q^2 \ov 6} - i \epsilon} \over {\mu_q \over \sqrt{6}} \pm k} \ .
 \ee
Note that the $1/\sqrt{6}$ factor multiplying $\mu_q$ in~\eqref{roep}
has the same origin as the one appearing in~\eqref{epeo}.

\section{Properties of spectral functions} \label{sec:3}

\subsection{General behavior}

We now describe the properties of $G_R$ obtained by solving~\eqref{sor1}. First note that by
taking $k \to - k$ the equations for $\xi_\pm$ exchange with each other, leading to
 \be \label{ne3}
 G_{22} (\om, k) = G_{11} (\om,-k) \ .
  \ee
Similarly by taking $q \to -q, \om \to -\om$ we find that
\be
G_{11}
(\om, k; -q) = - G_{22}^* (-\om, k;q) \ .
 \ee
So it is enough to restrict to positive $k$ and $q$. One can also check that as $|\om|,|k| \gg
\mu_q$, both components reduce to those of the vacuum. When $m=0$, by dividing the equation for
$\xi_+$ in~\eqref{sor1} by $\xi_+^2$, we find that  $\xi_- = - {1 \ov \xi_+}$, which implies
that
 \be \label{ne2}
G_{22} (\om, k) = - {1 \ov G_{11} (\om, k)}, \qquad m=0 \ .
\ee
Combining~\eqref{ne3} and~\eqref{ne2} we thus conclude that at $k=0$,
 \be \label{eep}
 G_{11} (\om,k=0) = G_{22} (\om,k=0) = i, \quad m=0 \ .
 \ee

\begin{figure}[h]
\vskip-0.15in
\hfill\hskip-1in \includegraphics[scale=0.39]{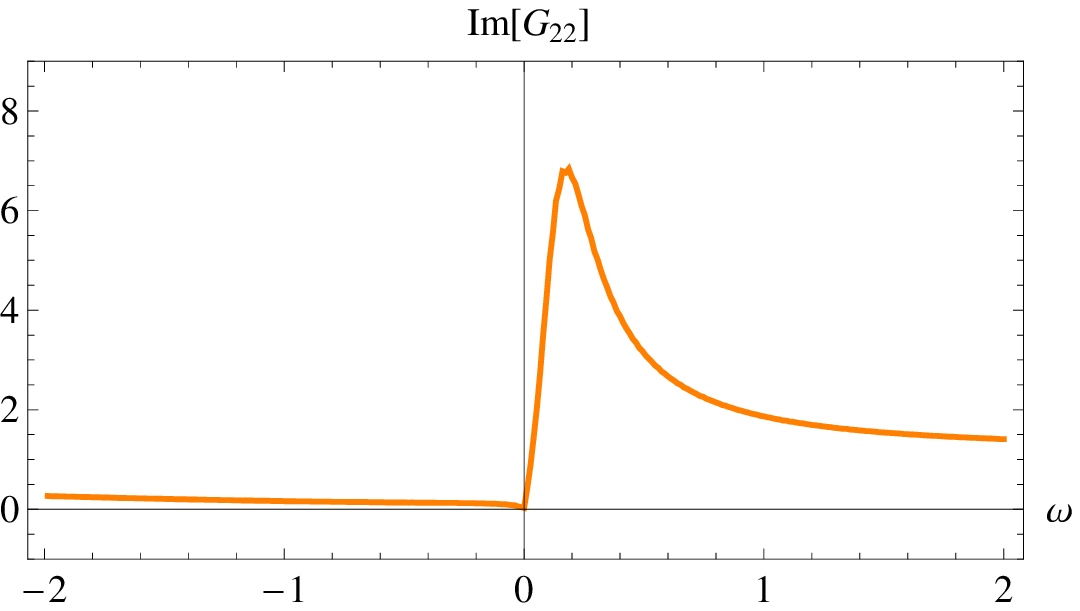}
\includegraphics[scale=0.4]{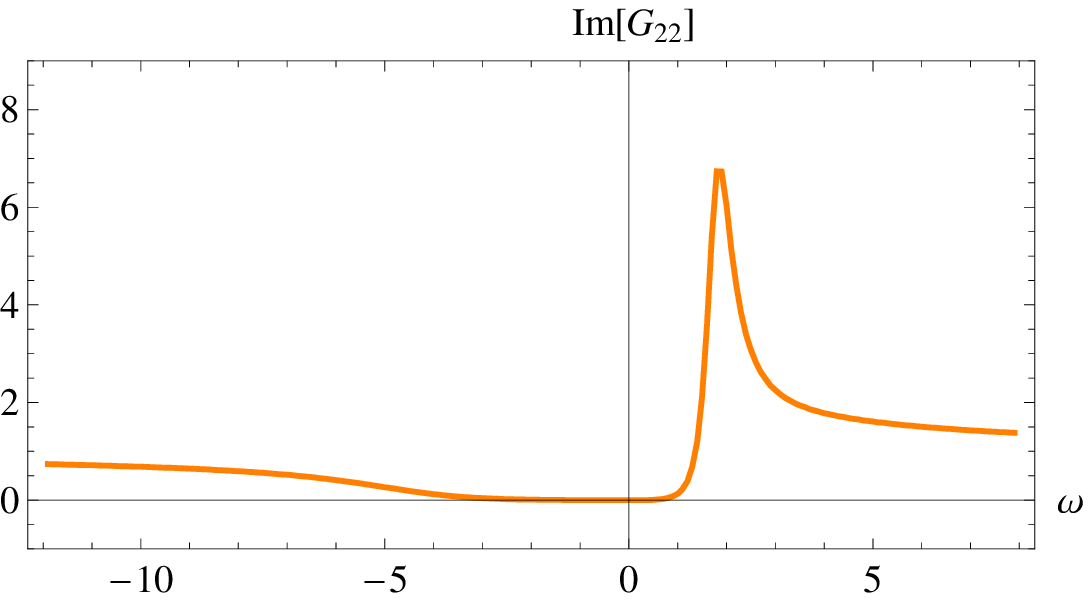}\hskip-0.1in\hfill
\caption{\label{fig:Sp} Spectral function $\Im G_{22}(\omega)$ at $k=1.2 < \mu_q$ (left plot) and
$k=3.0 > \mu_q$ (right plot) for $m=0$ and $q=1\; (\mu_q = \sqrt{3})$. The function asymptotes to $1$ as
$|\om| \to \infty$ as in the vacuum~\eqref{vac}. Right plot: The onset of the finite peak  at
$\om \approx 1.2 \approx k - \mu_q$ roughly corresponds to the location of divergence at $\om = k$ in the vacuum~\eqref{vac}.
The function is roughly zero between $\om \in (-k - \mu_q, k-\mu_q)$,
as it is in vacuum.
Left plot: The deviation from the vacuum behavior becomes significant.
}
\vskip-0.05in
\end{figure}

\begin{figure}[h] \begin{center}
\includegraphics[scale=0.45,angle=270]{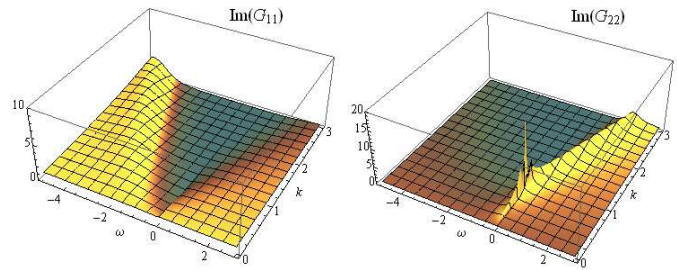}
\caption{\label{fig:three_d} 3d plots of $\Im G_{11}(\omega, k)$ and $\Im G_{22}(\omega, k)$ for
$m=0$ and $q=1 \, (\mu_q = \sqrt{3})$. In the right plot the ridge at $k \gg \mu_q$ corresponds to the
smoothed-out peaks at finite density of the divergence at $\om = k$ in the vacuum. As one decreases $k$ to a value $k_F \approx 0.92 < \mu_q$, the ridge
in $\Im G_{22}$ develops into an (infinitely) sharp peak indicative of a Fermi surface.
}
\end{center}
\end{figure}

Further study of $G_R$ is possible by numerically solving \eqref{sor1}. We will first consider
$T=0$ and will mostly discuss the massless case. The mass dependence will be discussed briefly
at the end. There are several consistency checks on our numerics. Firstly, $\Im G_{11}$ and $\Im
G_{22}$ are both positive, which is a requirement of unitarity since the diagonal components are
proportional to spectral densities. For a fixed large $k \gg \mu_q$, $\Im G_{11}$ has a
linearly-dispersing constant-height peak at $\om + \mu_q \approx -k$ and $\Im G_{22}$ has a peak
at $\om + \mu_q \approx k$, while both components are roughly zero in the region $\om + \mu_q
\in (-k, k)$~(see figure~\ref{fig:Sp} and~\ref{fig:three_d}). This recovers the behavior in the
vacuum, which is given by~\cite{Henningson:1998cd,Mueck:1998iz}
 \be \label{vac}
G_{11} = -\sqrt{k - (\om + i \ep) \ov k + (\om +
i\ep)}, \quad G_{22} = \sqrt{k + (\om + i \ep) \ov k - (\om + i\ep)}
 \ee
with now the divergences at $\om = \pm k$ smoothed out into finite size peaks.

\subsection{Fermi surface}

As one decreases $k$ to $\mu_q$ and smaller, the behavior of $G_R$ deviates significantly from
that of the vacuum. For definiteness, let us now focus on $q=1$ (with $\mu_q = \sqrt{3}$). In
this case the finite peak of $\Im G_{22}$ in the large $k$ region develops into a sharp
quasi-particle-like peak near $k_F = 0.918528499(1)$ (see figure~\ref{fig:three_d}). The
behavior of $\Im G_{22}$ in the region of small $k_\perp\equiv k-k_F$ and $\om$ can be
summarized as follows:

\ben

\begin{figure}[h]
\begin{center}
  \includegraphics[totalheight=5.5cm,origin=c,angle=0]{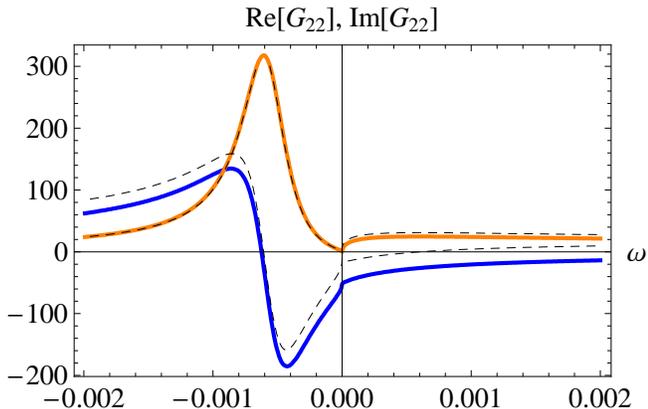}
\caption{\label{fig:lkf}  $\Re G_{22}(\omega)$ (blue) and $\Im G_{22}(\omega)$ (orange) at
$k=0.90 < k_F$. In $\Im G_{22}$, at $\omega<0$ there is a quasi-particle-like peak; for
$\omega>0$, there is a much smaller `bump'. As $k$ approaches $k_F$, the peak and the bump
approach $\om =0$ and their heights approach infinity. The dashed lines are the real and
imaginary parts of the fit function~\eqref{rro1}.  Although the real part slightly deviates from
the fit, there is a qualitative match. }
\end{center}
\end{figure}

\item For $k_\perp < 0$, we find a sharp quasi-particle-like peak in the region $\om < 0$ and a small
``bump'' (with a broad maximum) in the region $\om > 0$ (see figure~\ref{fig:lkf}). This appears
to indicate that there is a quasi-particle-like pole in the left quadrant of the lower-half
complex $\om$-plane. As $k_\perp \to 0_-$, both the peak and the maximum of the bump approach
$\om =0$, their heights go to infinity, and their widths go to zero. By carefully examining when
the peak and the bump meet we are able to determine the accuracy of $k_F= 0.918528499(1)$ to
10th digit.

\begin{figure}[h]
\begin{center}
  \includegraphics[totalheight=8cm,origin=c,angle=0]{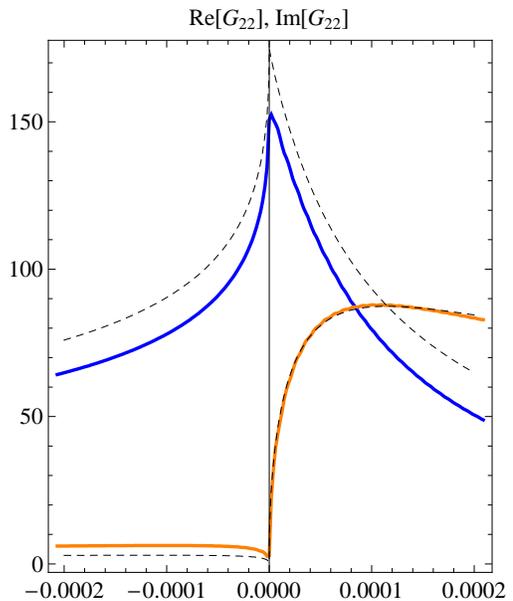}
\caption{\label{fig:rkf}  $\Re G_{22}(\omega)$ (blue) and $\Im G_{22}(\omega)$ (orange) at $k
=0.925 > k_F$. One finds a ``bump'' at $\omega> 0$ and a much smaller ``bump'' at $\om < 0$. As
$k_\perp$ approaches $0_+$, both bumps approach $\om =0$ and their heights approach infinity.
The dashed lines are the real and imaginary parts of the fit function~\eqref{Bee}. The fit is
not so good for $\om < 0$, though the qualitative trend matches. }
\end{center}
\end{figure}

\item For $k_\perp>0$, one does {\it not} see a sharp peak along real $\om$-axis for either sign of
$\om$. Instead one finds a ``bump'' (with a broad maximum) on the $\om > 0$ side and a smaller
bump on the $\om < 0$ side. See figure~\ref{fig:rkf}. In the limit $k_\perp \to 0_+$, both bumps
approach $\om =0$ and their heights go into infinity.

\begin{figure}[h]
\begin{center}
  \includegraphics[totalheight=3.1cm,origin=c,angle=0]{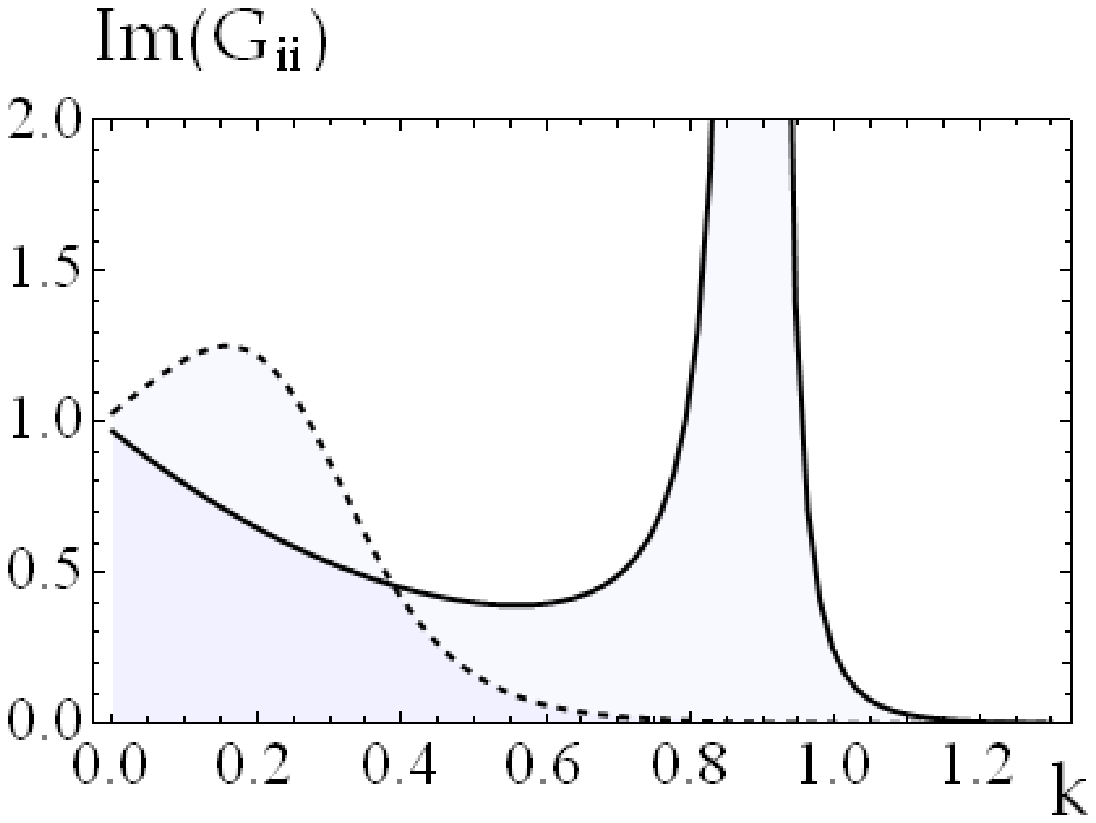}
  \includegraphics[totalheight=3.1cm,origin=c,angle=0]{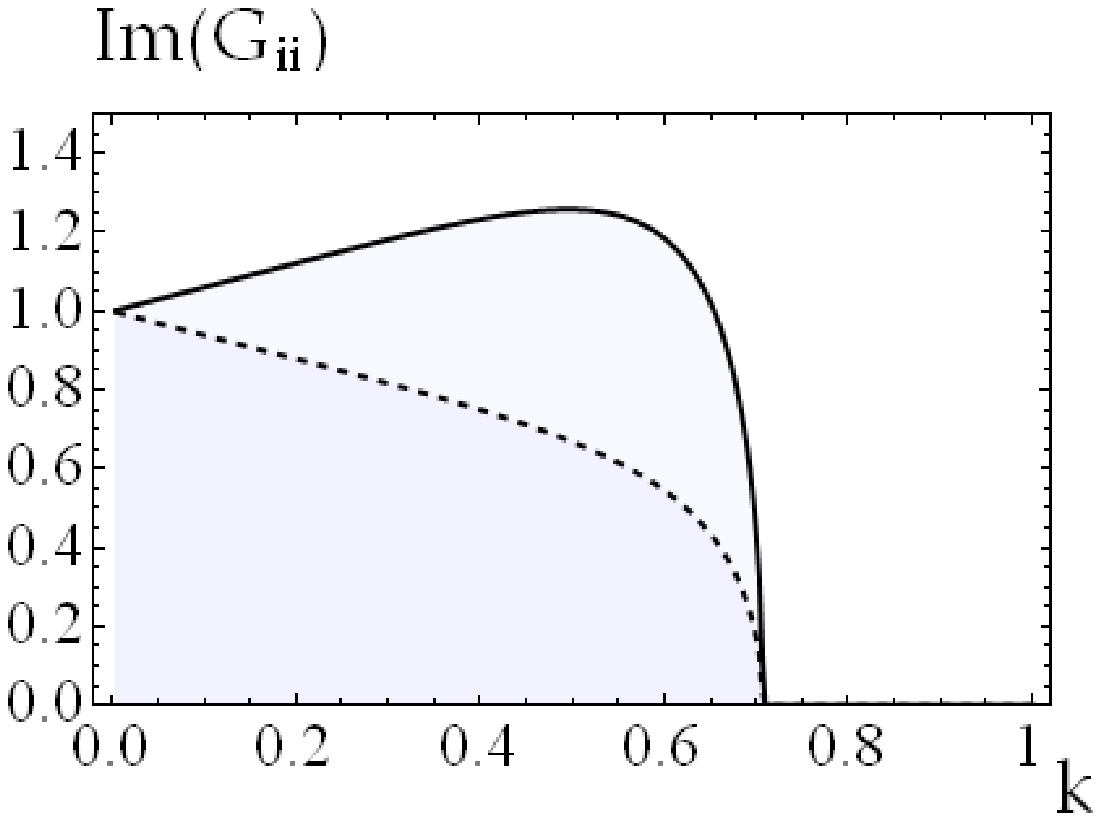}
\caption{\label{fig:fixw} Left: Plots of $\Im G_{11} (k)$ (dashed line) and  $\Im G_{22} (k)$ as
a function of $k$ at $\om = -0.001$ ($m=0$ and $q=1$). A sharp peak in $\Im G_{22} (k)$ is
clearly visible near $k_F \approx 0.9185$. The height of the peak is finite. In the limit $\om
\to 0_-$, the height of the peak goes to infinity and the location of the peak approaches $k_F$
from left. Right: Plots of $\Im G_{11} (k)$ (dashed line) and  $\Im G_{22} (k)$ as a function of
$k$ at $\om =0$. For $\omega=0$, both functions become identically zero in the region $k >
{\mu_q \ov \sqrt 6} = {1 \ov \sqrt{2}}$. Since $k_F > 1/\sqrt{2}$, at $\om =0$, $\Im G_{22}$ is
identically zero around $k_F$. }
\end{center}
\end{figure}

\item The quasi-particle-like peak and various bumps can also be studied by plotting $\Im G_{22} (k,
\om)$ as a function of $k$ for a given $\om$ (see the left panel of figure~\ref{fig:fixw} for a
plot at $\om =-0.001$). In the limit $\om \to 0_-$, the height of the peak goes to infinity with
its width  going to zero. At exactly $\om =0$, however, the functions $\Im G_{11}$ and $\Im
G_{22}$ become identically zero for $k > {\mu_q \ov \sqrt{6}} = {1 \ov \sqrt{2}}$ (see the right
panel of figure~\ref{fig:fixw}). This behavior can be understood from~\eqref{roep}
and~\eqref{sor1} as follows. For $k \geq {\mu_q \ov \sqrt{6}}$ (at $m=0$), the boundary
conditions~\eqref{roep} become real and since~\eqref{sor1} are real equations, $\Im G_{ii}
(\om=0, k)$ are then identically zero in this region. Note that $k_F > {1 \ov \sqrt{2}} $.

\een

\begin{figure}[h]
\hfill\hskip-1in \includegraphics[scale=0.70]{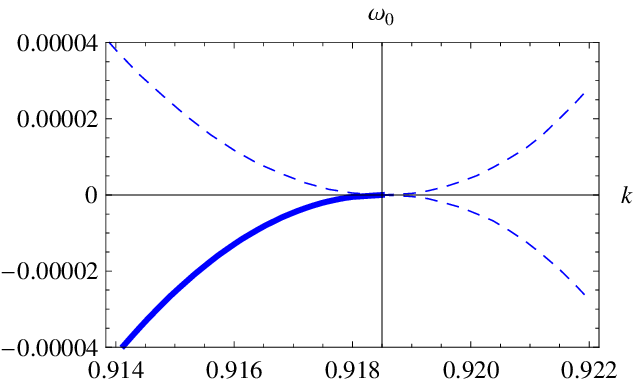}
\includegraphics[scale=0.36]{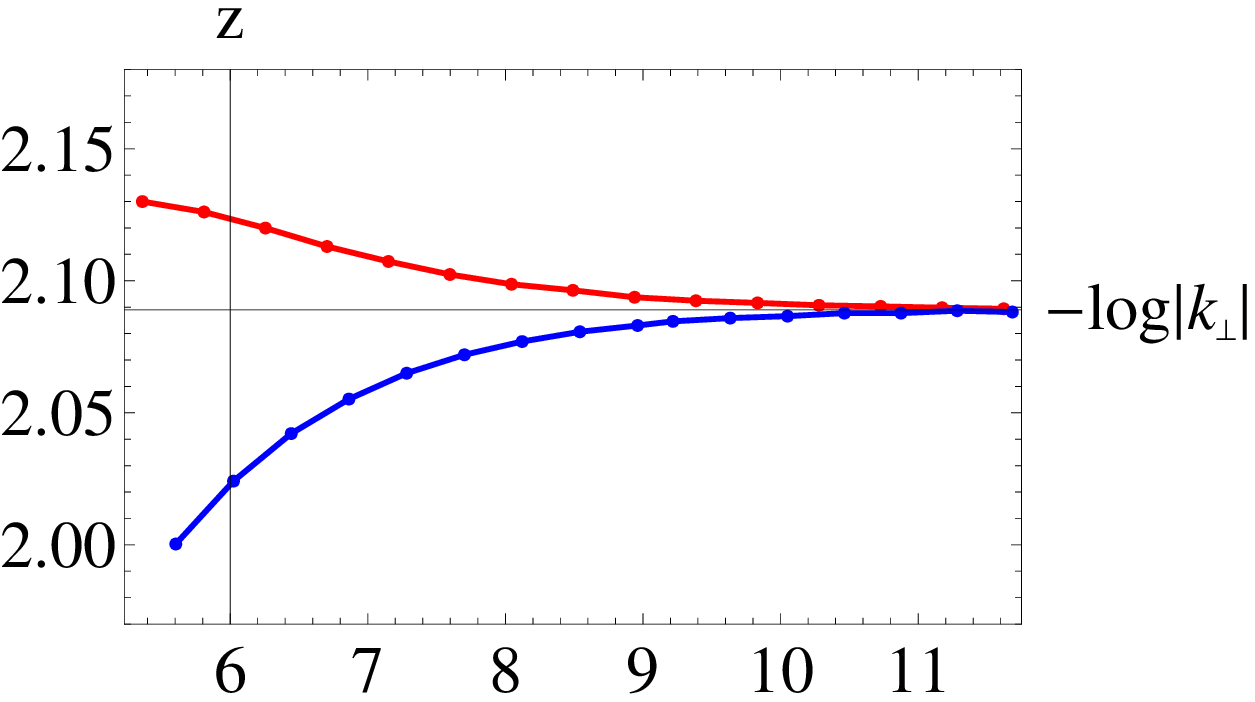}
\hskip-0.1in\hfill \caption{\label{fig:lim}  Left plot: Dispersion relation $\omega_*(k) \sim
k_\perp^z$ around $k_F$. The dashed lines indicate the $\omega$-values of the maxima of the
bumps. The solid line shows the dispersion of the quasi-particle peak. Right plot: Convergence
of the $z$ scaling exponent for $k>k_F$ (red) and $k<k_F$ (blue). The horizontal axis is the
natural log of $k_\perp$. } \vskip-0.05in
\end{figure}

Denoting the location of the maximum of the quasi-particle-like peak as $\om_* (k_\perp)$ we
find that $\om_* (k_\perp)$ scales with $k_\perp \to 0_-$ as
 \be \label{ooe1}
 \om_* (k_\perp) \sim k_\perp^z, \qquad z = 2.09 \pm 0.01
 \ee
and the height of $\Im G_{22}$ at the maximum scales as
 \be
 \Im G_{22} (\om_* (k_\perp), k_\perp) \sim k_\perp^{-\al}, \qquad   \al = 1.00 \pm 0.01 \
 \ee
(see figure~\ref{fig:lim}). One finds exactly the same scaling behavior also for the maxima of
the other three ``bumps''. This strongly suggests that in the limit of small $k_\perp$ and $\om$
$\Im G_{22} (\om, k_\perp)$ has the following scaling form
 \be \label{ppe}
 \Im G_{22} (\lam^z \om, \lam k_\perp) = \lam^{-\al} \Im G_{22} ( \om,  k_\perp)
 \ee
with  the scaling exponents $\al,z$ given by
 \be \label{ppe1}
 z = 2.09 \pm 0.01, \qquad \al = 1.00 \pm 0.01 \ .
 \ee

The scaling behavior~\eqref{ooe1}--\eqref{ppe1} suggests an underlying sharp Fermi surface with
Fermi momentum $k_F$. It is, however, not of the form corresponding to a {\it Landau} Fermi
liquid which would have exponents $z = \al =1$. It is an example of the more general scaling
behavior discussed recently by Senthil~\cite{senthil1, senthil2} for a critical Fermi surface
occurring at a continuous metal-insulator transition\footnote{For early work,
see~\cite{cha1,cha2}.}. The system also has a rather curious particle-hole asymmetry; the
quasi-particle-like peak at $k_\perp<0$ morphs into a ``bump'' at $k_\perp > 0$ as the Fermi
surface is crossed (compare feature 1 and 2 of previous page). The fact that $\mu_q > k_F$
suggests that the system has repulsive interactions.

For a given $\om \neq 0$, $G_{ii} (k, \om)$ are non-singular for any value of $k$ including
$k_F$, while for a given value of $k$,  $G_{ii} (k, \om)$ is continuous but non-smooth at $\om
=0$.\footnote{From numerical calculation, it does appear that the functions become smoother
for $k \gg \mu_q$.} This non-smooth behavior at $\om=0$ for momenta away from the Fermi momentum
is puzzling and it would be nice to understand its physical interpretation better. From the
gravity side, this is related to the aforementioned singular behavior of
equations~\eqref{set1}--\eqref{set2} and~\eqref{sor1} near $\omega \to 0$, as discussed
around~\eqref{roep}. This can be further attributed to the existence of an AdS$_2$
region~\eqref{epeo} in the bulk geometry at zero temperature.

Further support for the scaling behavior~\eqref{ppe} near the Fermi surface can be obtained by
fitting the whole curves of $G_{22} (\om)$ (rather than just the behavior near the maxima) for
different $k_\perp$ by a scaling function which is analytic in the upper half
$\om$-plane\footnote{We would like to thank S.~Sachdev for a discussion of possible
subtleties.}:

\ben

\item For $k_\perp < 0$, $G_{22}$ can be fitted\footnote{The function fits well not only along the real $\om$-axis, but also in the upper
half plane. There are numerical instabilities at zero temperature in the lower half complex
$\om$-plane and we have not been able to perform a direct fit there.} for both signs of $\om$, by (see figure~\ref{fig:lkf})
 \be \label{rro1}
 G_{22} (\om, k_\perp) \approx {c_0 (-k_\perp)^{-\al} \ov \log \le({-\om \ov c_1 (-k_\perp)^z} \ri) - i \ga}
 \ee
where $\ga \approx 0.34$ and $c_0, c_1$ are positive constants (in the scaling region). The
above function has a pole in the lower half $\om$-plane at
 \be
 \om_c = - c_1 (-k_\perp)^z e^{i \ga} \ .
 \ee
As $k_\perp \to 0_-$, $\om_c$ approaches to the origin of the complex plane along a straight
line which has an angle $\ga - \pi$ with respect to the positive real axis. Since $\Re \om_c$
gives the location $\om_* (k_\perp)$ of the peak and $-\Im \om_c$ gives the width $\Ga$ of the
peak, for~\eqref{rro1},
 \be
 \Ga = \tan \ga \, |\om_* (k_\perp)| \ .
 \ee
This linear dependence of the width on $\omega_\star$\footnote{Recall that for a Landau
quasi-particle, $\Ga \sim \om_*^2$.} is reminiscent of the behavior in \eg \cite{varma}.

\item For $k_\perp>0$, we have not found a good global fit for both signs of $\om$. A reasonable fit for the imaginary part is

 \be \label{Bee}
 G_{22} (\om, k_\perp) \approx  {a_0 k_\perp^{-\al} \over a_1 - i \le({|\omega| \ov k_\perp^z} \ri)^{\al \ov z}}~ ;
\ee
where $a_0$ and $a_1$ are positive constants which take different values for $\om < 0$ and $\om > 0$\footnote{Again we are handicapped by a numerical instability the
lower half $\om$-plane which prevents a fit directly around the singularities in the lower half
plane.}.
\een
It is important to note that the functions~\eqref{rro1} and~\eqref{Bee} are only best numerical
fits we could find and should not be taken too seriously as the ``genuine'' functions which
describe the system. Both are consistent with the requirement from figure~\ref{fig:fixw} that
$\Im G_{22} (k_\perp) =0$ at $\om =0$. The different fit functions for $k_\perp< 0$ and $k_\perp
> 0$ may reflect the ``particle-hole asymmetry'' discussed earlier. Also note that for a nonzero
$k_\perp$ both~\eqref{rro1} and~\eqref{Bee} indicate a branch point singularity at $\om =0$, but
have different $k_\perp \to 0$ limits. The behavior of $\Im G_{22} (\om)$ at exactly $k= k_F$ is
not completely clear to us at the moment.

\subsection{Discrete scale invariance}

\begin{figure}[h]
\hfill\hskip-1in
\includegraphics[scale=0.50,origin=c,angle=0]{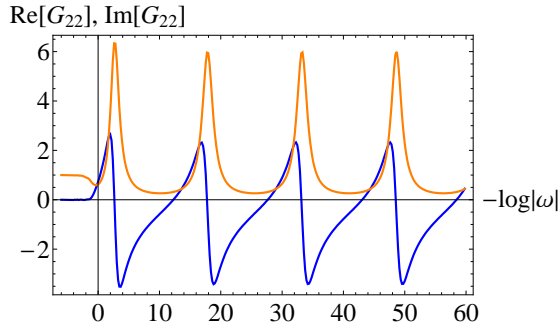}
\caption{\label{fig:conv}
   Both $\Re G_{22}(\omega, k=0.5)$ (blue curve) and $\Im G_{22} (\om,k=0.5)$ (orange curve) are periodic in $\log \om$ as $\om \to 0$.
   The period appears to be given by
   $ \Delta(\log \omega) \approx  \frac{\pi\sqrt{6}}{\sqrt{ \mu_q^2/6 - (k^2 +m^2) }}$.  This formula was guessed
   based on the behavior of the solution in the $AdS_2$ region;
   the formula is confirmed by the numerics.}
\end{figure}

In the region $k < {\mu_q \ov \sqrt{6}}$, where $\Im G_{ii} (\om=0, k)$ are nonzero (see
figure~\ref{fig:fixw}), a new phenomenon occurs in the $\om \to 0$ limit. One finds that $\Im
G_{ii} (\om, k)$ become oscillatory with oscillatory peaks periodic in $\log |\om|$ with
constant heights, see fig.~\ref{fig:conv}. More explicitly we find
 \be \label{dsw}
 G_{ii} (\om, k) = G_{ii} (\om e^{n \xi (k)} , k), \qquad n \in \ZZ, \quad \om \to 0
 \ee
where $\xi (k)$ is a ($k$-dependent) positive constant. In other words, $G_{ii}$ is invariant
under a discrete scaling in $\om$. $\xi (k)$ appears to decrease with $k$ and approaches a
constant in the limit $k \to 0$. In the limit $k \to {\mu_q \ov \sqrt{6}}$, $\xi(k)$ approaches
infinity. The height of the oscillatory peaks also decreases with $k$, approaching zero as $k
\to 0$ (where the whole function approaches unity) and a finite constant as $k \to {\mu_q \ov
\sqrt{6}}$. It would be interesting to understand whether~\eqref{dsw} is associated with some
kind of complex scaling exponents. Below we will  refer to the region $k < {\mu_q \ov \sqrt{6}}$
as the oscillatory region. Note that the oscillatory region appears to be the counterpart for
fermions of the unstable region for a charged boson, where the corresponding bosonic modes have
complex energies and want to condense. In the fermion case, the oscillatory region does not
appear to indicate an instability, \eg there is no singularity in the upper half of the complex
$\om$-plane.

\subsection{Finite temperature}

\begin{figure}[h]
\begin{center}
  \includegraphics[totalheight=2.91cm,origin=c,angle=0]{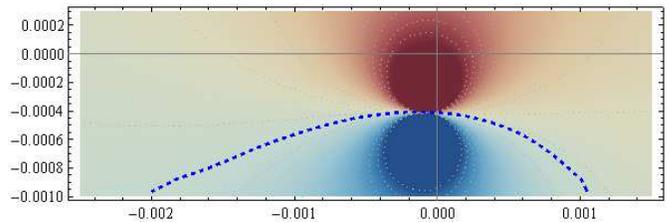}
\caption{\label{complex_omega}The complex omega plane for $T=4.13 \times 10^{-4}$: now the
quasi-particle pole is finite distance below the real $\om$-axis. The dashed line indicates the trajectory of the pole between $k=0.87$(left)$\ldots 0.93$(right). The closest distance to the
real axis is equal to the temperature $T$ (up to 1\% accuracy). There is a numerical instability for $ \Im \omega < -\pi T$ which can also be seen directly from the wave equation. We leave it for future work to explore this part of the lower half plane. Also shown is the density plot for
$\Im G_{22}(\omega)$ at $k = 0.90$, where the corresponding pole is closest to the real axis.}
\end{center}
\end{figure}

Turning on a small temperature $T$ appears to smooth everything out. There is no longer a sharp
Fermi surface, \ie\ there no longer exists a sharp momentum at which $\Im G_{22}$ becomes
singular for any real $\om$ and $k$. Going to the lower half $\om$-plane, one finds that all the
singularities are a finite distance away from the real axis, with the closest distance given by
$T$ which happens at $k \approx 0.90$ (see figure~\ref{complex_omega})\footnote{Similar results
have also been obtained by Carlos Fuertes. We thank Carlos Fuertes and Subir Sachdev for
communicating the results to us.}. This behavior is different from the Fermi liquid where the
width is quadratic in temperature. Note that for a given small $k_\perp < 0$, as one turns on
the temperature, the corresponding quasi-particle-like pole in the complex $\om$-plane appears
to move down and to the right. It is also interesting to note that at finite $T$, there are now
quasi-particle-like poles for momenta $k>k_F$. Perhaps they are generated from the branch point
at $T=0$.

At finite $T$, the functions $\Im G_{ii}$ become smooth at $\om =0$ and in the oscillatory
region there are only a finite number of oscillations as the $\om \to 0$ limit is approached.

\subsection{Charge dependence}

\begin{figure}[h]
\begin{center}
\hskip -0.2cm
\includegraphics[scale=0.42,angle=270]{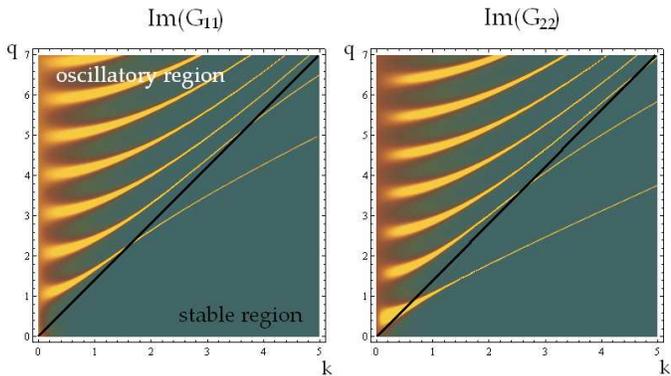}
  \caption{\label{fig:qvsk} Density plot of $\Im G_{ij}(k, q)_{\omega=-0.001}$ with $k \in [0, 5]$ and $q \in [0,7]$ at $T= 2.76\times 10^{-6}$. A negligible temperature was turned on in order to increase the speed of the computation. The results were not affected by this. The orange lines are locations where the functions become very large.
Also note that the width of the peaks in $\Im G(k)$ decreases quickly as one moves towards
larger charge. The black line is $k=\mu_q/\sqrt{6}$ to the left of which is the oscillatory region.
  }
\end{center}\end{figure}

When we increase (decrease) $q$ to be greater (smaller) than $1$, the Fermi momentum $k_F$
increases (decreases) with $q$ approximately linearly. As $q$ is further increased, new branches
of Fermi surfaces appear. These features can be seen in figure~\ref{fig:qvsk}, which gives the
density plots of $\Im G_{11}$ and $\Im G_{22}$ in the $q-k$ plane at a fixed value of
$\om=-0.001$.

We have sampled the exponents $z,\alpha$ for a few other values of $q$ for the lowest branch of
fermi surface in $\Im G_{22}$, \eg for $q=0.6$, $z \approx 5.32, \; \alpha \approx 1.00$, and
for $q=1.2$, $z \approx 1.53, \; \al \approx 1.00$. Compared to the values for $q=1$ described
earlier, it then appears that $z$ decreases rapidly with increasing $q$, while $\al=1$ is
independent of $q$. Note that in~\cite{senthil1} it was argued that $z \geq \al$ and $z \geq 1$.
Thus it could be that $z$ will asymptote to $1$ for larger values of $q$.\footnote{At larger
values of $q$ the  convergence of the exponents becomes slower; we leave this for future work.
Also note that the value $\al=1$ is special according to the scaling theory of~\cite{senthil1}.}
We also find that the constant $\ga$ in~\eqref{rro1} appears to decrease rapidly with $q$. Thus
it seems likely that as $q$ is increased, the non-Fermi liquid will become more like a Landau
Fermi liquid. Given that $k_F$ increases with $q$, this is reminiscent of asymptotic freedom in
high-density QCD.

The $q-k$ space in figure~\ref{fig:qvsk} is separated into two regions by the (black) $k={\mu_q
\ov \sqrt{6}}$ line. In the region to right (stable region), the locations of the quasi-particle
lines (\ie\ orange lines in~figure~\ref{fig:qvsk}) stabilize in the limit $\omega \rightarrow 0$
and indicate locations of Fermi surfaces. The region to the left is the oscillatory region
discussed earlier, where the log-periodic oscillatory behavior is reflected in a downward motion
of the orange lines as $|\om|$ is decreased; they seem to become infinitely dense in the limit
$\omega \rightarrow 0$. Also recall that in the oscillatory region, the heights of the peaks
remain finite in the $\om \to 0$ limit.

\begin{figure}[h] \begin{center}
\includegraphics[totalheight=9cm,angle=270]{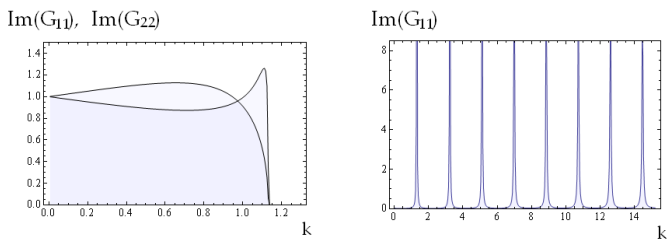}
  \caption{\label{envelope} (i) $\Im G_{ij}(k)_{\omega=0}$ with $q=1.6$. Near $k=\mu_q/\sqrt{6}$, a bump is seen in $\Im G_{22}(k)$.
  At slightly higher value of the charge, the Fermi surface crosses the boundary of the oscillatory region and this bump becomes a peak. \ (ii)  $\Im G_{11}(q, k)_{\omega=0}$ along the
  $\mu_q=\sqrt{6}k$ line. The spacing between the peaks is constant, $\Delta k \sim \sqrt{7/2} (\pm 1\%)$. }
\end{center}\end{figure}

As one decreases $q$, a Fermi surface line in figure~\ref{fig:qvsk} will intersect the line $k =
{\mu_q \ov \sqrt{6}}$, disappear into the oscillatory region, and lose its status as a Fermi
surface. Thus the behavior of $\Im G_{ii}$ in the oscillatory region is strongly correlated with
the sprouting of new branches of Fermi surface as $q$ is varied (see figure~\ref{envelope}).

We have also studied other values of the mass (with $m<\half$) at $q=1$, and find the Fermi
momentum $k_F$ decreases linearly with increasing mass. At finite mass, the oscillatory region
is now given by $k^2 + m^2 < {\mu_q^2 \ov 6}$ (see~\eqref{roep}). We find that for $q=1$, the
Fermi surface disappears into the oscillatory region at roughly $m \approx 0.4$. For $m > {\mu_q
\ov \sqrt{6}}$, the oscillatory region disappears; we expect the Fermi surface will also
disappear beyond this value if not before.

\subsection{Summary}

To conclude this section, let us summarize the main features of the spectral functions which we
have observed. The system we study is conformally invariant at zero density. Turning on a finite
charge density breaks Lorentz and scale invariance. The energy scale of the problem is
controlled by the chemical potential $\mu$ which for a charge $q$ particle becomes $\mu_q = \mu
q$. At $q=1$ the spectral functions also exhibit two other interesting scales. The first is the
Fermi momentum $k_F < \mu_q$ around which we observe a quasi-particle-like peak which suggests
an underlying Fermi surface. The scaling behavior and the particle-hole asymmetry around the
Fermi surface indicate that this is a non-Fermi liquid. The other scale is $k_S \equiv {\mu_q
\ov \sqrt{6}}$, which lies below $k_F$. We find that for $k < k_S$, the spectral functions have
log-periodic oscillatory behavior near $\om =0$, which indicates some underlying discrete scale
invariance. At larger values of $q$ new scales corresponding to more than branches of Fermi
surface also appear.

It is important to emphasize that the scaling behavior observed here is not related to the scale
invariance of the vacuum, which is broken by the nonzero charge density. It is emergent, arising
as a consequence of the collective behavior of many particles. Note that both the scaling
behavior around the Fermi surface and the discrete scale invariance involve the small $\om$
limit, which on the gravity side can be attributed to the AdS$_2$ region in the near horizon
geometry of the black hole at zero temperature. It may be possible that this emergent scaling
behavior can be understood from the AdS$_2$ region\footnote{Work in progress with T.~Faulkner
\cite{newpaper}.}.

\section{Discussion} \label{sec:4}

Finally, we discuss some caveats and possible interpretation of the results. While the black
hole geometry~\eqref{bhmetric} is by itself thermodynamically and perturbatively stable, when it
is embedded into a specific gravity theory, new instabilities may occur. For example, it is
possible for charged (bulk) scalars to condense, spontaneously breaking the $U(1)$
symmetry~\cite{Gubser:2008px,Hartnoll:2008vx}. This happens if the bulk spectrum includes a
charged scalar of sufficiently large charge or sufficiently small mass~\cite{Denef:2009tp}. The
boundary theories considered in~\cite{Denef:2009tp} all contain such scalars including the
$\sN=8$ M2 brane theory and ABJM theory. It would be very interesting to understand how the
condensate affects the Fermi surfaces and scaling behavior observed here, and how generic the
existence of such scalars is.

The black hole solution~\eqref{bhmetric} has a finite entropy at zero temperature, and thus
describes an ensemble of an exponentially large number of states. Given that the solution is not
supersymmetric, beyond the gravity approximation likely these states are energetically
closely-spaced, rather than exactly degenerate. This ``frustrated property'' is shared by many
known models of spin liquids. The behavior described above then reflects average behavior of a
large number of states rather than that of a single ground state.

A rather mysterious feature of our results is that different probe operators appear to find
different Fermi surface structure which depends on (and only on) their charges and operator
dimensions. One possible explanation is as follows\footnote{We would like to thank T. Senthil
for long and very instructive discussions about this.}. Let us look at the OPE of {\it e.g.} the
first component $\sO_1$ of a fermionic operator $\sO$, which has the schematic form
\be
\label{OPE}
\sO_1 (\ep)^\dagger \sO_1 (0) \sim {1 \over \ep^{2 \Delta} } + {c(\De) \ov N^2} { q J^0 \over \ep^{ 2\Delta -2 } } + \cdots,
\ee
where $J^0$ is the zero component of global $U(1)$ current.
\eqref{OPE} implies that the density for $\sO_1$ can be written as
\be \label{erp}
n_{\sO_1} = \vev{\sO_1 (\ep)^\dagger \sO_1 (0)} \sim {1 \over \ep^{2 \Delta} } + {c(\De) \ov N^2} { q \vev{J^0} \over \ep^{ 2\Delta -2 } } + \cdots \qquad \ep \to 0
\ee
where $\ep$ should be considered as a short-distance cutoff. The first term on RHS
of~\eqref{erp} is the standard piece due to vacuum fluctuations, which can be subtracted.  In a
state of  finite density, the second term induces a density for $\sO_1$, which is proportional
to $q$, the background charge density and depends on the UV cutoff through the dimension of the
operator. Since the Fermi surface involves modes with wavelengths of order $ k_F^{-1}$, which is
parametrically distinct from the short wavelength of the modes contributing the UV divergence,
we expect the fermionic density which is responsible for the Fermi surface should be insensitive
to the UV cutoff. The background charge density $\vev{J^0} \propto N^2$ induces a nonzero charge
density of order $O(N^0)$ for each charged operator, which depends on its charge and conformal
dimension. The induced density increases with its charge, which is consistent with our empirical
observation that $k_F$ increases with the charge of the bulk particle.

In the large $N$ limit, the effective interactions of $\sO$ with itself and other gauge
invariant operators are all suppressed by $1/N$. This may lead one to conclude that the
effective theory for $\sO$ should be a free Fermi gas, which would contradict our observed
scaling behavior near $k_F$. Note, however, the effective dynamics of $\sO$ can be different
from that of a free fermion, again due to large $N$ effects. To see this, let us look at the
current density from $\sO$, which can be schematically written as $j^\mu = \bar \sO \ga^\mu
\sO$. The fluctuations of $j^\mu$, which can be read from its connected two point functions, are
suppressed by $1/N^2$. Thus in the large $N$ limit, $j^\mu$ does not fluctuate. One can try to
model this by coupling a free fermion to a gauge field, which acts as a Lagrange multiplier
suppressing the fluctuations of the associated current. When coupling a Fermi liquid to a
dynamical gauge field, it is well known (see {\it e.g.}~\cite{Holstein:1973zz, reizer,
Polchinski:1993ii, Halperin:1992mh, Nayak:1993uh, Nayak:1994ng,altshuler-1994,
Boyanovsky:2000bc, Ipp:2003cj, Schafer:2004zf}) that long-range magnetic interactions result in
a non-Fermi liquid, which appears to be consistent with our picture. It would be desirable to
make this argument more precise. Note that the particle-hole asymmetry is not seen in previously
known models. The above suggestion does not preclude the existence of some fundamental non-Fermi
liquid structure from which the behavior of probe fermionic operators could be derived.

The fact that the induced charge density for each probe operator is of order $O(1)$ also implies
that their contributions to the transport of the system are not visible at leading order in the
large $N$ expansion. Indeed, to leading order in $N$ none of the observables like specific heat,
conductivity, entanglement entropy can depend on the charge or dimension of probe spinor fields.
However, if there exists a fundamental non-Fermi liquid structure, some effects might still be
visible at leading order. We will leave this for future study.

Finally, as indicated earlier, the near horizon AdS$_2$ region appears to play a role for the
appearance of both the log-periodic behavior in the oscillatory region and the Fermi-surfaces.
Clearly it would be interesting to have a better understanding of the CFT interpretation of the AdS$_2$ region.

\vspace{0.2in}   \centerline{\bf{Acknowledgements}} \vspace{0.2in} We thank T.~Faulkner,
C.~Fuertes, S.~Hartnoll, N.~Iqbal, V.~Khodel, P.~Lee, M.~Mulligan, K.~Rajagopal, S.~Sachdev,
S.~Shenker, B.~Swingle, G.~Volovik, B.~Zwiebach and in particular T.~Senthil for valuable
discussions and encouragement. Work supported in part by funds provided by the U.S. Department
of Energy (D.O.E.) under cooperative research agreement DE-FG0205ER41360 and the OJI program.

\vspace{1in}

\vfill\eject

\bibliography{draft}

\begin{thebibliography}{40}
\expandafter\ifx\csname natexlab\endcsname\relax\def\natexlab#1{#1}\fi
\expandafter\ifx\csname bibnamefont\endcsname\relax
  \def\bibnamefont#1{#1}\fi
\expandafter\ifx\csname bibfnamefont\endcsname\relax
  \def\bibfnamefont#1{#1}\fi
\expandafter\ifx\csname citenamefont\endcsname\relax
  \def\citenamefont#1{#1}\fi
\expandafter\ifx\csname url\endcsname\relax
  \def\url#1{\texttt{#1}}\fi
\expandafter\ifx\csname urlprefix\endcsname\relax\def\urlprefix{URL }\fi
\providecommand{\bibinfo}[2]{#2}
\providecommand{\eprint}[2][]{\url{#2}}

\bibitem[{\citenamefont{Anderson}(1990)}]{anderson}
\bibinfo{author}{\bibfnamefont{P.~W.} \bibnamefont{Anderson}},
  \bibinfo{journal}{Phys. Rev. Lett.} \textbf{\bibinfo{volume}{64}},
  \bibinfo{pages}{1839} (\bibinfo{year}{1990}).

\bibitem[{\citenamefont{Varma et~al.}(1989)\citenamefont{Varma, Littlewood,
  Schmitt-Rink, Abrahams, and Ruckenstein}}]{varma}
\bibinfo{author}{\bibfnamefont{C.~M.} \bibnamefont{Varma}},
  \bibinfo{author}{\bibfnamefont{P.~B.} \bibnamefont{Littlewood}},
  \bibinfo{author}{\bibfnamefont{S.}~\bibnamefont{Schmitt-Rink}},
  \bibinfo{author}{\bibfnamefont{E.}~\bibnamefont{Abrahams}}, \bibnamefont{and}
  \bibinfo{author}{\bibfnamefont{A.~E.} \bibnamefont{Ruckenstein}},
  \bibinfo{journal}{Phys. Rev. Lett.} \textbf{\bibinfo{volume}{63}},
  \bibinfo{pages}{1996} (\bibinfo{year}{1989}).

\bibitem[{\citenamefont{Varma et~al.}(2002)\citenamefont{Varma, Nussinov, and
  {van Saarloos}}}]{varmarev}
\bibinfo{author}{\bibfnamefont{C.~M.} \bibnamefont{Varma}},
  \bibinfo{author}{\bibfnamefont{Z.}~\bibnamefont{Nussinov}}, \bibnamefont{and}
  \bibinfo{author}{\bibfnamefont{W.}~\bibnamefont{{van Saarloos}}},
  \bibinfo{journal}{Physics Reports} \textbf{\bibinfo{volume}{361}},
  \bibinfo{pages}{267} (\bibinfo{year}{2002}), \eprint{cond-mat/0103393}.

\bibitem[{\citenamefont{{Benfatto} and {Gallavotti}}(1990)}]{Benfatto}
\bibinfo{author}{\bibfnamefont{G.}~\bibnamefont{{Benfatto}}} \bibnamefont{and}
  \bibinfo{author}{\bibfnamefont{G.}~\bibnamefont{{Gallavotti}}},
  \bibinfo{journal}{Journal of Statistical Physics}
  \textbf{\bibinfo{volume}{59}}, \bibinfo{pages}{541} (\bibinfo{year}{1990}).

\bibitem[{\citenamefont{Shankar}(1991)}]{ShankarA}
\bibinfo{author}{\bibfnamefont{R.}~\bibnamefont{Shankar}},
  \bibinfo{journal}{Physica} \textbf{\bibinfo{volume}{A177}},
  \bibinfo{pages}{530} (\bibinfo{year}{1991}).

\bibitem[{\citenamefont{Polchinski}(1992)}]{Polchinski:1992ed}
\bibinfo{author}{\bibfnamefont{J.}~\bibnamefont{Polchinski}}
  (\bibinfo{year}{1992}), \eprint{hep-th/9210046}.

\bibitem[{\citenamefont{Shankar}(1994)}]{Shankar:1993pf}
\bibinfo{author}{\bibfnamefont{R.}~\bibnamefont{Shankar}},
  \bibinfo{journal}{Rev. Mod. Phys.} \textbf{\bibinfo{volume}{66}},
  \bibinfo{pages}{129} (\bibinfo{year}{1994}).

\bibitem[{\citenamefont{Maldacena}(1998)}]{Maldacena:1997re}
\bibinfo{author}{\bibfnamefont{J.~M.} \bibnamefont{Maldacena}},
  \bibinfo{journal}{Adv. Theor. Math. Phys.} \textbf{\bibinfo{volume}{2}},
  \bibinfo{pages}{231} (\bibinfo{year}{1998}), \eprint{hep-th/9711200}.

\bibitem[{\citenamefont{Gubser et~al.}(1998)\citenamefont{Gubser, Klebanov, and
  Polyakov}}]{Gubser:1998bc}
\bibinfo{author}{\bibfnamefont{S.~S.} \bibnamefont{Gubser}},
  \bibinfo{author}{\bibfnamefont{I.~R.} \bibnamefont{Klebanov}},
  \bibnamefont{and} \bibinfo{author}{\bibfnamefont{A.~M.}
  \bibnamefont{Polyakov}}, \bibinfo{journal}{Phys. Lett.}
  \textbf{\bibinfo{volume}{B428}}, \bibinfo{pages}{105} (\bibinfo{year}{1998}),
  \eprint{hep-th/9802109}.

\bibitem[{\citenamefont{Witten}(1998)}]{Witten:1998zw}
\bibinfo{author}{\bibfnamefont{E.}~\bibnamefont{Witten}},
  \bibinfo{journal}{Adv. Theor. Math. Phys.} \textbf{\bibinfo{volume}{2}},
  \bibinfo{pages}{505} (\bibinfo{year}{1998}), \eprint{hep-th/9803131}.

\bibitem[{\citenamefont{Bagger and Lambert}(2008)}]{Bagger:2007vi}
\bibinfo{author}{\bibfnamefont{J.}~\bibnamefont{Bagger}} \bibnamefont{and}
  \bibinfo{author}{\bibfnamefont{N.}~\bibnamefont{Lambert}},
  \bibinfo{journal}{JHEP} \textbf{\bibinfo{volume}{02}}, \bibinfo{pages}{105}
  (\bibinfo{year}{2008}), \eprint{0712.3738}.

\bibitem[{\citenamefont{Gustavsson}(2009)}]{Gustavsson:2007vu}
\bibinfo{author}{\bibfnamefont{A.}~\bibnamefont{Gustavsson}},
  \bibinfo{journal}{Nucl. Phys.} \textbf{\bibinfo{volume}{B811}},
  \bibinfo{pages}{66} (\bibinfo{year}{2009}), \eprint{0709.1260}.

\bibitem[{\citenamefont{Aharony et~al.}(2008)\citenamefont{Aharony, Bergman,
  Jafferis, and Maldacena}}]{Aharony:2008ug}
\bibinfo{author}{\bibfnamefont{O.}~\bibnamefont{Aharony}},
  \bibinfo{author}{\bibfnamefont{O.}~\bibnamefont{Bergman}},
  \bibinfo{author}{\bibfnamefont{D.~L.} \bibnamefont{Jafferis}},
  \bibnamefont{and}
  \bibinfo{author}{\bibfnamefont{J.}~\bibnamefont{Maldacena}},
  \bibinfo{journal}{JHEP} \textbf{\bibinfo{volume}{10}}, \bibinfo{pages}{091}
  (\bibinfo{year}{2008}), \eprint{0806.1218}.

\bibitem[{\citenamefont{Lee}(2008)}]{Lee:2008xf}
\bibinfo{author}{\bibfnamefont{S.-S.} \bibnamefont{Lee}}
  (\bibinfo{year}{2008}), \eprint{0809.3402}.

\bibitem[{\citenamefont{Son and Starinets}(2002)}]{Son:2002sd}
\bibinfo{author}{\bibfnamefont{D.~T.} \bibnamefont{Son}} \bibnamefont{and}
  \bibinfo{author}{\bibfnamefont{A.~O.} \bibnamefont{Starinets}},
  \bibinfo{journal}{JHEP} \textbf{\bibinfo{volume}{09}}, \bibinfo{pages}{042}
  (\bibinfo{year}{2002}), \eprint{hep-th/0205051}.

\bibitem[{\citenamefont{Iqbal and Liu}(2009{\natexlab{a}})}]{Iqbal:2008by}
\bibinfo{author}{\bibfnamefont{N.}~\bibnamefont{Iqbal}} \bibnamefont{and}
  \bibinfo{author}{\bibfnamefont{H.}~\bibnamefont{Liu}},
  \bibinfo{journal}{Phys. Rev.} \textbf{\bibinfo{volume}{D79}},
  \bibinfo{pages}{025023} (\bibinfo{year}{2009}{\natexlab{a}}),
  \eprint{0809.3808}.

\bibitem[{\citenamefont{Iqbal and Liu}(2009{\natexlab{b}})}]{iqbalN}
\bibinfo{author}{\bibfnamefont{N.}~\bibnamefont{Iqbal}} \bibnamefont{and}
  \bibinfo{author}{\bibfnamefont{H.}~\bibnamefont{Liu}}
  (\bibinfo{year}{2009}{\natexlab{b}}), \eprint{0903.2596}.

\bibitem[{\citenamefont{Chamblin et~al.}(1999)\citenamefont{Chamblin, Emparan,
  Johnson, and Myers}}]{Chamblin:1999tk}
\bibinfo{author}{\bibfnamefont{A.}~\bibnamefont{Chamblin}},
  \bibinfo{author}{\bibfnamefont{R.}~\bibnamefont{Emparan}},
  \bibinfo{author}{\bibfnamefont{C.~V.} \bibnamefont{Johnson}},
  \bibnamefont{and} \bibinfo{author}{\bibfnamefont{R.~C.} \bibnamefont{Myers}},
  \bibinfo{journal}{Phys. Rev.} \textbf{\bibinfo{volume}{D60}},
  \bibinfo{pages}{064018} (\bibinfo{year}{1999}), \eprint{hep-th/9902170}.

\bibitem[{\citenamefont{Romans}(1992)}]{Romans:1991nq}
\bibinfo{author}{\bibfnamefont{L.~J.} \bibnamefont{Romans}},
  \bibinfo{journal}{Nucl. Phys.} \textbf{\bibinfo{volume}{B383}},
  \bibinfo{pages}{395} (\bibinfo{year}{1992}), \eprint{hep-th/9203018}.

\bibitem[{\citenamefont{Gauntlett and Varela}(2007)}]{Gauntlett:2007ma}
\bibinfo{author}{\bibfnamefont{J.~P.} \bibnamefont{Gauntlett}}
  \bibnamefont{and} \bibinfo{author}{\bibfnamefont{O.}~\bibnamefont{Varela}},
  \bibinfo{journal}{Phys. Rev.} \textbf{\bibinfo{volume}{D76}},
  \bibinfo{pages}{126007} (\bibinfo{year}{2007}), \eprint{0707.2315}.

\bibitem[{\citenamefont{Henningson and Sfetsos}(1998)}]{Henningson:1998cd}
\bibinfo{author}{\bibfnamefont{M.}~\bibnamefont{Henningson}} \bibnamefont{and}
  \bibinfo{author}{\bibfnamefont{K.}~\bibnamefont{Sfetsos}},
  \bibinfo{journal}{Phys. Lett.} \textbf{\bibinfo{volume}{B431}},
  \bibinfo{pages}{63} (\bibinfo{year}{1998}), \eprint{hep-th/9803251}.

\bibitem[{\citenamefont{Mueck and Viswanathan}(1998)}]{Mueck:1998iz}
\bibinfo{author}{\bibfnamefont{W.}~\bibnamefont{Mueck}} \bibnamefont{and}
  \bibinfo{author}{\bibfnamefont{K.~S.} \bibnamefont{Viswanathan}},
  \bibinfo{journal}{Phys. Rev.} \textbf{\bibinfo{volume}{D58}},
  \bibinfo{pages}{106006} (\bibinfo{year}{1998}), \eprint{hep-th/9805145}.

\bibitem[{\citenamefont{Senthil}(2008{\natexlab{a}})}]{senthil1}
\bibinfo{author}{\bibfnamefont{T.}~\bibnamefont{Senthil}}
  (\bibinfo{year}{2008}{\natexlab{a}}), \eprint{0803.4009}.

\bibitem[{\citenamefont{Senthil}(2008{\natexlab{b}})}]{senthil2}
\bibinfo{author}{\bibfnamefont{T.}~\bibnamefont{Senthil}}
  (\bibinfo{year}{2008}{\natexlab{b}}), \eprint{0804.1555}.

\bibitem[{\citenamefont{Yin and Chakravarty}(1996)}]{cha1}
\bibinfo{author}{\bibfnamefont{L.}~\bibnamefont{Yin}} \bibnamefont{and}
  \bibinfo{author}{\bibfnamefont{S.}~\bibnamefont{Chakravarty}},
  \bibinfo{journal}{Int. J. Mod. Phys.} \textbf{\bibinfo{volume}{B10}},
  \bibinfo{pages}{805} (\bibinfo{year}{1996}).

\bibitem[{\citenamefont{Chakravarty et~al.}(1998)\citenamefont{Chakravarty,
  Yin, and Abrahams}}]{cha2}
\bibinfo{author}{\bibfnamefont{S.}~\bibnamefont{Chakravarty}},
  \bibinfo{author}{\bibfnamefont{L.}~\bibnamefont{Yin}}, \bibnamefont{and}
  \bibinfo{author}{\bibfnamefont{E.}~\bibnamefont{Abrahams}},
  \bibinfo{journal}{Phys. Rev. B} \textbf{\bibinfo{volume}{58}},
  \bibinfo{pages}{R559} (\bibinfo{year}{1998}).

\bibitem[{\citenamefont{Faulkner et~al.}(2009)\citenamefont{Faulkner, Liu,
  McGreevy, and Vegh}}]{newpaper}
\bibinfo{author}{\bibfnamefont{T.}~\bibnamefont{Faulkner}},
  \bibinfo{author}{\bibfnamefont{H.}~\bibnamefont{Liu}},
  \bibinfo{author}{\bibfnamefont{J.}~\bibnamefont{McGreevy}}, \bibnamefont{and}
  \bibinfo{author}{\bibfnamefont{D.}~\bibnamefont{Vegh}}
  (\bibinfo{year}{2009}), \eprint{0907.2694}.

\bibitem[{\citenamefont{Gubser}(2008)}]{Gubser:2008px}
\bibinfo{author}{\bibfnamefont{S.~S.} \bibnamefont{Gubser}},
  \bibinfo{journal}{Phys. Rev.} \textbf{\bibinfo{volume}{D78}},
  \bibinfo{pages}{065034} (\bibinfo{year}{2008}), \eprint{0801.2977}.

\bibitem[{\citenamefont{Hartnoll et~al.}(2008)\citenamefont{Hartnoll, Herzog,
  and Horowitz}}]{Hartnoll:2008vx}
\bibinfo{author}{\bibfnamefont{S.~A.} \bibnamefont{Hartnoll}},
  \bibinfo{author}{\bibfnamefont{C.~P.} \bibnamefont{Herzog}},
  \bibnamefont{and} \bibinfo{author}{\bibfnamefont{G.~T.}
  \bibnamefont{Horowitz}}, \bibinfo{journal}{Phys. Rev. Lett.}
  \textbf{\bibinfo{volume}{101}}, \bibinfo{pages}{031601}
  (\bibinfo{year}{2008}), \eprint{0803.3295}.

\bibitem[{\citenamefont{Denef and Hartnoll}(2009)}]{Denef:2009tp}
\bibinfo{author}{\bibfnamefont{F.}~\bibnamefont{Denef}} \bibnamefont{and}
  \bibinfo{author}{\bibfnamefont{S.~A.} \bibnamefont{Hartnoll}}
  (\bibinfo{year}{2009}), \eprint{0901.1160}.

\bibitem[{\citenamefont{Holstein et~al.}(1973)\citenamefont{Holstein, Norton,
  and Pincus}}]{Holstein:1973zz}
\bibinfo{author}{\bibfnamefont{T.}~\bibnamefont{Holstein}},
  \bibinfo{author}{\bibfnamefont{R.~E.} \bibnamefont{Norton}},
  \bibnamefont{and} \bibinfo{author}{\bibfnamefont{P.}~\bibnamefont{Pincus}},
  \bibinfo{journal}{Phys. Rev.} \textbf{\bibinfo{volume}{B8}},
  \bibinfo{pages}{2649} (\bibinfo{year}{1973}).

\bibitem[{\citenamefont{Reizer}(1989)}]{reizer}
\bibinfo{author}{\bibfnamefont{M.~Y.} \bibnamefont{Reizer}},
  \bibinfo{journal}{Phys. Rev. B} \textbf{\bibinfo{volume}{40}},
  \bibinfo{pages}{11571} (\bibinfo{year}{1989}).

\bibitem[{\citenamefont{Polchinski}(1994)}]{Polchinski:1993ii}
\bibinfo{author}{\bibfnamefont{J.}~\bibnamefont{Polchinski}},
  \bibinfo{journal}{Nucl. Phys.} \textbf{\bibinfo{volume}{B422}},
  \bibinfo{pages}{617} (\bibinfo{year}{1994}).

\bibitem[{\citenamefont{Halperin et~al.}(1993)\citenamefont{Halperin, Lee, and
  Read}}]{Halperin:1992mh}
\bibinfo{author}{\bibfnamefont{B.~I.} \bibnamefont{Halperin}},
  \bibinfo{author}{\bibfnamefont{P.~A.} \bibnamefont{Lee}}, \bibnamefont{and}
  \bibinfo{author}{\bibfnamefont{N.}~\bibnamefont{Read}},
  \bibinfo{journal}{Phys. Rev.} \textbf{\bibinfo{volume}{B47}},
  \bibinfo{pages}{7312} (\bibinfo{year}{1993}).

\bibitem[{\citenamefont{Nayak and Wilczek}(1994{\natexlab{a}})}]{Nayak:1993uh}
\bibinfo{author}{\bibfnamefont{C.}~\bibnamefont{Nayak}} \bibnamefont{and}
  \bibinfo{author}{\bibfnamefont{F.}~\bibnamefont{Wilczek}},
  \bibinfo{journal}{Nucl. Phys.} \textbf{\bibinfo{volume}{B417}},
  \bibinfo{pages}{359} (\bibinfo{year}{1994}{\natexlab{a}}),
  \eprint{cond-mat/9312086}.

\bibitem[{\citenamefont{Nayak and Wilczek}(1994{\natexlab{b}})}]{Nayak:1994ng}
\bibinfo{author}{\bibfnamefont{C.}~\bibnamefont{Nayak}} \bibnamefont{and}
  \bibinfo{author}{\bibfnamefont{F.}~\bibnamefont{Wilczek}},
  \bibinfo{journal}{Nucl. Phys.} \textbf{\bibinfo{volume}{B430}},
  \bibinfo{pages}{534} (\bibinfo{year}{1994}{\natexlab{b}}).

\bibitem[{\citenamefont{Altshuler et~al.}(1994)\citenamefont{Altshuler, Ioffe,
  and Millis}}]{altshuler-1994}
\bibinfo{author}{\bibfnamefont{B.~L.} \bibnamefont{Altshuler}},
  \bibinfo{author}{\bibfnamefont{L.~B.} \bibnamefont{Ioffe}}, \bibnamefont{and}
  \bibinfo{author}{\bibfnamefont{A.~J.} \bibnamefont{Millis}}
  (\bibinfo{year}{1994}), \eprint{cond-mat/9406024}.

\bibitem[{\citenamefont{Boyanovsky and de~Vega}(2001)}]{Boyanovsky:2000bc}
\bibinfo{author}{\bibfnamefont{D.}~\bibnamefont{Boyanovsky}} \bibnamefont{and}
  \bibinfo{author}{\bibfnamefont{H.~J.} \bibnamefont{de~Vega}},
  \bibinfo{journal}{Phys. Rev.} \textbf{\bibinfo{volume}{D63}},
  \bibinfo{pages}{034016} (\bibinfo{year}{2001}), \eprint{hep-ph/0009172}.

\bibitem[{\citenamefont{Ipp et~al.}(2004)\citenamefont{Ipp, Gerhold, and
  Rebhan}}]{Ipp:2003cj}
\bibinfo{author}{\bibfnamefont{A.}~\bibnamefont{Ipp}},
  \bibinfo{author}{\bibfnamefont{A.}~\bibnamefont{Gerhold}}, \bibnamefont{and}
  \bibinfo{author}{\bibfnamefont{A.}~\bibnamefont{Rebhan}},
  \bibinfo{journal}{Phys. Rev.} \textbf{\bibinfo{volume}{D69}},
  \bibinfo{pages}{011901} (\bibinfo{year}{2004}), \eprint{hep-ph/0309019}.

\bibitem[{\citenamefont{Schafer and Schwenzer}(2004)}]{Schafer:2004zf}
\bibinfo{author}{\bibfnamefont{T.}~\bibnamefont{Schafer}} \bibnamefont{and}
  \bibinfo{author}{\bibfnamefont{K.}~\bibnamefont{Schwenzer}},
  \bibinfo{journal}{Phys. Rev.} \textbf{\bibinfo{volume}{D70}},
  \bibinfo{pages}{054007} (\bibinfo{year}{2004}), \eprint{hep-ph/0405053}.

\end{thebibliography}

\end{document}